\documentclass[a4paper]{article}
\usepackage{amssymb}
\usepackage{bm}
\usepackage{graphicx}
\usepackage{amsmath}
\begin{document}
\title{Particle-Field Theory and Its Relativistic Generalization II ( Relativistic Generalization of Micro Harmonic Oscillator and Hydrogen Atom )}
\author{{ Fatemeh Ahmadi$^1$\thanks{e-mail: f.ahmadi@bzte.ac.ir; fatemehs.ahmadis@gmail.com;
} ,  \,\,\,Afshin Shafiee$^{2,3}$\thanks{ Corresponding
author, e-mail:
shafiee@sharif.edu}
}\\[0.4cm]
{ \em $^{1}$ \small Department of  Engineering Sciences and Physics }\\
{\em  \small Buein Zahra Technical University, Ghazvin, Iran.}\\
{ \em $^{2}$ \small Research Group on Foundations of Quantum Theory and Information,} \\
{\em \small Department of Chemistry, Sharif University of Technology,}\\
{\em \small  P.O. Box 11365-9516, Tehran, Iran}\\
{\em  $^{3}$  \small School of Physics, Institute for Research in Fundamental Sciences}\\
{\em \small  (IPM), P.O. Box 19395-5531,
Tehran, Iran}}
\date{}
\maketitle
\begin{abstract}
As a serious attempt for constructing a new foundation for describing micro-entities from a causal standpoint,
it was explained before in \cite{shafiee, shafiee2, shafiee3} that by unifying the concepts of information, matter and energy, each micro-entity is assumed
to be composed of a probability field joined to a particle called a particle-field or PF system. The relativistic generalization of  this theory
and  its invariance under Lorentz transformation has been proved.\\
\indent
In this essay,  based on the relativistic generalization of Schr\"{o}dinger equation  derived in \cite{shafiee4}, we solve  the relativistic Schr\"{o}dinger  equation for relativistic micro-harmonic oscillator to find its energy. Also we obtain the energy spectrum of Hydrogen atom that is the main purpose of this paper.  We see that the result is completely consistent with
the relativistic correction to the Hydrogen's energy in first-order perturbation theory.
\end{abstract}
\section{Introduction}
Quantum theory is the theoretical basis of modern physics that explains the nature and behavior of matter and energy on the atomic and subatomic level. It is undoubtedly one of the most important and experimentally accurate scientific theories in the history of science. It  continues to yield novel and unexpected results, in technology as well as in all scientific fields, including physics, biology, chemistry, and so on. However, in spite of all  empirical and mathematical success of quantum mechanics, due to its broad and ambiguous conceptual framework, until now,  all attempts to reach a satisfactory understanding of its meaning have been remained unpalatable. One can never expect an understanding of quantum mechanics that is similar in clarity and intelligibility to the one provided by other domains of physics. Its problems are of a fundamental nature, or they are more fundamental than  other cases.
Although  refined theoretical arguments and new experimental techniques have produced a substantial advance in the last recent decades,
 the famous sentence of Feynman, who said "nobody really understand quantum mechanics"\cite{Belal} is still marked.\\
 \indent
 We have attempted to construct a new foundation for describing micro-events from a deterministic causal standpoint, in which a micro-entity is supposed  to be an allied particle-field system, instead of composing of a particle and (or) a field (wave) \cite{shafiee,shafiee2,shafiee3}.
It has been explained in the first essay of this series that in the microworld, one encounters an unified concept of information,
matter and energy \cite{shafiee}.
In this new approach, the principles of realism and causality based on the classic-like equations of motion are presumed.
A particle-field system is not composed of a particle and a wave. Instead, it is a unified system for which the particle and the wave notions are only abstract
constructions without real manifestation.\\
\indent
Here, one may pose the question that what the differences are between this approach and Bohmian \cite{bohm,holland} account for a micro-system.
The point is that, a system is neither a particle nor a wave, not also a combination of these two entities. It is a totality of both wave and particle notions, so that one can imagine it as a field that enfolds a particle. We abstract the notions particle and field, the PF system to describe more elaborately. So, it seems that particle and field construct the PF system and when the energy of the field approaches zero, the classical particle appears. Yet, in reality, there are no distinct entities such as the particle and the field. Only when the PF system loses its all holistic nature, it reduces to a known classical particle.
 Thus, it looks like we have two different energies, one for the particle and the other for the field and the latter causes the quantum behavior of the system.\\
 \indent
 This important feature of a PF system enables one, e.g., to show why the squared modulus of the wave function behaves like a probability density in spatial
 coordinates. While, in Bohmian theory Born postulate is accepted a \emph{priori}.\\
\indent
Considering a PF system allows one to explain the origin of the Schr\"{o}dinger equation, since here we assume that the underlying dynamics of a supposed field is influenced by an oscillatory force which
could be approximated to a harmonic one in the first order [1]. Taking into account anharmonic effects, one can obtain non-linear forms of the Schr\"{o}dinger equation. In addition, from a fundamental point of view, these two theories explain bizarre quantum phenomena like the measurement problem, tunneling effect and double-slit experiment in completely different
directions. The interested reader can follow the corresponding fashions of explanation in each model in \cite{shafiee, shafiee2, shafiee3}
 and \cite{bohm, holland}.\\
\indent
Moreover, one of the most important matters  is that the PF theory is not in contradiction with Special Relativity
in its origin.
 The Lorentz-invariant forms of equations in PF theory to obtain  relativistic Schr\"{o}dinger equation has been presented \cite{shafiee4}.\\
\indent
Here, we are going to solve the relativistic Schr\"{o}dinger equation found in \cite{shafiee4} for relativistic micro-harmonic oscillator and relativistic Hydrogen atom to find their
energy spectrum.\\
\indent
The paper is organized as follows: In section 2, we review the basic elements of the PF
theory for a one-particle one-dimensional microsystem and its relativistic generalization.
In section 3, we solve the relativistic Schr\"{o}dinger  equation for  micro-harmonic
oscillator and in section 4, we find the energy of Hydrogen atom and show that it is consistent with
the result of relativistic correction discussed in text book \cite{Griffits}.  In section 5, the whole content of our paper is discussed and concluded.
%
\section{Review of Basic Elements}
In this section, we give a brief review of basic elements of the PF theory and its relativistic generalization.
 More details are available in \cite{shafiee, shafiee2, shafiee3, shafiee4}.\\
 \indent
 For a one-dimensional, one-particle microsystem, three physical entities are introduced:\\
\indent
1. A particle with mass $m$ and position $x(t)$ whose dynamics is given by the Newton's second law:
\begin{equation}
m\frac{d^{2}x(t)}{dt^{2}}=f_{P},
\end{equation}
where $f_{P}$ is the force defined for the particle. For the conservative forces, the particle possesses a conserved energy $E_{P}=V_{P}+K_{P}$, where $K_{P}=\frac{p^{2}_{P}}{2m}$ is the kinetic energy and $p_{P}$ is the linear momentum of the particle.\\
\indent
2. Like the particle aspect of the PF system, there is a field denoted by $X(x(t),t)$  with velocity
 $v_{F}=|\frac{dX}{dt}|=|\dot{X}|$ along the positive direction of $x$, where
\begin{equation}\label{}
\dot{X}=\left(\frac{\partial X}{\partial x}\right)v_{P}+\left(\frac{\partial X}{\partial t}\right),
\end{equation}
and $v_{P}$ is the velocity of the particle along the same direction. The amplitude of the field has a dimension of length.
Similar to the particle, we assume that the field obeys a Newton-like dynamics too in the same direction,
\begin{equation}\label{a34}
m\frac{d\dot{X}}{dt}=f_{F},
\end{equation}
where $f_{F}$  is the force the field is subjected to. If the particle is subjected to a conservative force $f_{P}$, we shall consider $X=\chi(x(t))$.
 Then, one can show that
%
\begin{equation}\label{a35}
f_{F}=mv^{2}_{P}\frac{d|\chi^{\prime}|}{dx}+|\chi^{\prime}|f_{P}.
\end{equation}
From a physical point of view, the field $X$ merely enfolds the particle. It experiences its own mechanical-like force introduced as $f_{F}$ in (\ref{a34}). Although the presence  of the particle is essential for defining the force of the field. If there is no particle, there will not be any associated field too. The existence of the field depends on the existence of the particle, but the opposite is not true, because $X$ is a function of particle's position, not vice versa.\\
\indent
For a conservative field subjected to the force $f_{F}$ in (\ref{a35}), one can define the energy $E_{F}=V_{F}+K_{F}$ where $K_{F}=\frac{1}{2}mv^{2}_{F}=K_{P}|\chi^{\prime}|^{2}$. The kinetic energy of the field includes the kinetic energy of the particle. Here, one can't separate the meaning of  $K_{F}$ from $K_{P}$.\\
\indent
In the quantum domain,  the quantities $E_{P}$ and $E_{F}$ are not practically discernible, but the total energy $E=E_{P}+E_{F}$
is an observable property. One can write the total energy as:
\begin{eqnarray}\label{a36}
E&=&V_{P}+(E_{F}+\frac{p^{2}_{P}}{2m}),\nonumber \\
&=&V_{P}+\frac{p^{2}}{2m}
\end{eqnarray}
where $\frac{p^{2}}{2m}=(E_{F}+\frac{p^{2}_{P}}{2m})$, and $V_{P}$ is the particle's potential. \\
\indent
Unfortunately, the form of the force $f_{F}$ in (\ref{a35}) is complicated and unknown a \emph{priori}, and so it is not possible to obtain it
from (\ref{a34}) or (\ref{a35}). Accordingly, we postulate that for stationary states for which the energy is conserved, the form of
$X=\chi(x(t))$ could be obtained from the time-independent Schr\"{o}dinger equation:
\begin{equation}\label{a100}
\chi^{\prime \prime}=-k^{2}\chi
\end{equation}
where
$$k^{2}=\frac{p^{2}}{\hbar^{2}} =\frac{2m}{\hbar}(E-V_{p})$$
and
$$\chi^{\prime \prime}=\frac{d^{2}\chi(x)}{dx^{2}}$$
%
\indent
For stationary states in which $\chi$ is a real function, one can rewrite relation (\ref{a35}) as
\begin{equation}\label{a101}
f_{F}=-m\overline{w}^{2}\chi+f_{P}\chi^{\prime}
\end{equation}
in which we have used (\ref{a100}) and $\overline{w}^{2}=v_{P}^{2}k^{2}$. This shows that when $\chi$ is real and relying on
$x(t)$ only, the field-at least partly-experiences an oscillatory force. Here, $\overline{w}$ depends on $x(t)$ and the first term in
(\ref{a101}) does not actually describe an oscillating force, but has the same form.\\
3. Neither the particle, nor the field representation alone is adequate for explaining the physical behavior of a microsystem, comprehensively. What really gives us a thorough understanding of the nature of a quantum system is a holistic depiction of both particle and its associated field which we call  here a PF system. The kinetic energy of a PF system is proportional to
$K_{P}+K_{F}$, but its total energy is the same as $E$ in (\ref{a36}).
Let us define the kinetic energy of a PF system as $K_{PF}=\frac{1}{2}m\dot{q}^{2}$, where $q$ denotes the position of the PF and $\dot{q}$
is its velocity. Then, it is legitimate to suppose that $K_{PF}\propto K_{P}+K_{F}$, or
\begin{equation}\label{a361}
\dot{q}^{2}=g_{PF}^{2}(\dot{x}^{2}+|\dot{X}|^{2})
\end{equation}
where $g_{PF}$ is a proportionality factor and $\dot{x}=v_{P}$. For many problems, this factor is equal to one, but the non-oneness of its value in general is crucial in some other problems \cite{shafiee,shafiee2}.\\
\indent
%
%
From The above relation, one can obtain the trajectories of a PF system:
\begin{equation}\label{a38}
q(x,t)=g_{PF}\int dx\sqrt{\left(1+|\frac{dX(x,t)}{dx}|^{2}\right)}.
\end{equation}
\indent
The relation (\ref{a38}) shows that while we expect the particle to move along the infinitesimal displacement $dx$ in the $x$ direction, the displacement of the whole system is equal to $dq$, not $dx$. The difference here is due to the existence of the associated field which adds a new term, in addition to the direction the particle moves along. Hence, the PF system indeed keeps going through an integrated path determined by the whole action of the particle and its associated field.\\
\indent
Using the relation (\ref{a38}), one can obtain the finite displacement $q$ of a PF system in terms of the particle's location $x(t)$ and time, when the field $X(x,t)$ is known. Then, if the form of dependence of $x$ to $t$ is also known for a given physical problem, it is possible to write $q$ totally in terms of $t$. For stationary states, however, $q=q(x(t))$ and there is no explicit time-dependency. Therefore, one can see that the time variable could be kept concealed in equations of motions, so that the spatial direction $x$ would be sufficient for illustrating the behavior of $q$.\\
\indent
The dynamics of the PF system can also be described according to a Newtonian equation. So, we have
\begin{equation}\label{a1b}
m\frac{d^{2}q}{dt^{2}}=f_{PF},
\end{equation}
where $f_{PF}$ is the force the PF system is subjected to.\\
%
\indent
To show that this theory is consistent with the theory of special relativity, we need to find a unified concept of spacetime that is invariant under Lorentz transformation. Using  the definition of relativistic
kinetic energy of the stationary field and the PF system, we have found the relativistic generalization
of (\ref{a38}) as \cite{shafiee4},
\begin{equation}\label{a4}
\dot{q}=c\left( 1-\frac{1}{[(\gamma_{p}-1)(1+\chi^{\prime 2})+1]^{2}} \right)^{-\frac{1}{2}},
\end{equation}
where $c$ is speed of light, $\gamma_{P}=\frac{1}{\sqrt{1-\frac{v_{P}^{2}}{c^{2}}}}$ and $\chi^{\prime}=\frac{d\chi}{dx}$.  Using (\ref{a4}), then, we have shown that

\begin{equation}\label{a5}
ds^{2}=c^{2}dt^{2}-dq^{2}
\end{equation}
is invariant under Lorentz transformation \cite{shafiee4}.\\
\indent
Now, one can derive new form of the relativistic Schr\"{o}dinger equation.  The physical structure of the PF formalism which has constitutional similarities
to classical equations of motion permits us to derive a well-defined relativistic Schr\"{o}dinger equation
for stationary fields, regardless of the spin variable.\\
\indent
The dynamics of a stationary real field in one dimension (denoted by $\chi=\chi(x(t))$ in the relativistic regime can be represented as
\begin{equation}
\frac {d(m_{p}\dot{\chi})}{dt}=f_{rF},
\end{equation}
where $f_{rF}$ is the force defined for the field under the relativistic conditions  and  $m_{p}$ is the relativistic mass of
the particle:
\begin{equation}\label{a23}
m_{p}=\gamma_{p}m_{0};\hspace{1cm}\gamma_{p}=\left(1-\frac{v^{2}_{p}}{c^{2}}\right)^{-\frac{1}{2}},
\end{equation}
where $m_{0}$  is the rest mass, as before and $v_{P}$ is the velocity defined for the particle. The stationary field $\chi(x(t))$ does not explicitly depend on time. So, one can find out that
\begin{equation} \label{a24}
f_{rF}=f_{rP}\chi^{\prime}+\gamma_{P}m_{0}v^{2}_{P}\chi^{\prime\prime}
\end{equation}
where $\chi^{\prime}=\frac{d\chi}{dx}$ and $f_{rP}=m_{0}\dot{v}_{P}\gamma_{P}^{3}$ is the force exerted on the particle. It has been argued
that for stationary real fields, there exists an oscillating-like term in the force expression (denoted by the second term in (\ref{a24}), when $\gamma_{P} \rightarrow 1$) from which the non relativistic  time-independent Schr\"{o}dinger equation can be resulted
\cite{shafiee}.
Here, we suppose that the same situation holds true under the relativistic conditions. That is, for stationary real fields, we postulate
the following equality as a general rule:
\begin{equation}\label{a25}
-m_{P}\bar{w}^{2}\chi= \gamma_{P}m_{0}v^{2}_{P}\chi^{\prime\prime},
\end{equation}
where $m_{P}$ was defined in relation (\ref{a23}) and $\bar{w}^{2}=k^{2}v^{2}_{P}$. Here again, we define $k=\frac{p}{\hbar}$,
where p is the relativistic de Broglie momentum. From the relation  (\ref{a25}), it is
immediately concluded that
\begin{equation}\label{a26}
-\hbar^{2}\chi^{\prime\prime}=p^{2}\chi,
\end{equation}
which has the same form as the non-relativistic Schr\"{o}dinger equation. After some calculations to find an appropriate relation for $p^{2}$ in
relation (\ref{a26}), one can derive  relativistic Schr\"{o}dinger equations for the cases that the
potential energy of the  particle includes the relativistic mass as \cite{shafiee4}:
\begin{equation}\label{a32}
-\frac{\hbar^{2}}{2m_{0}}\chi^{\prime\prime}+\frac{1}{2}m_{0}c^{2}\chi=
\frac{E^{2}}{2m_{0}c^{2}}\left(1+\frac{V_{nrP}}{m_{0}c^{2}}\right)^{-2}\chi,
\end{equation}
and for potential energy that is independent of mass as  \cite{shafiee4}:
\begin{equation}\label{a33}
-\frac{\hbar^{2}}{2m_{0}}\chi^{\prime\prime}+\frac{1}{2}m_{0}c^{2}\chi=\frac{1}{2m_{0}c^{2}}(E-V_{nrP})^{2}\chi.
\end{equation}
%
In the following,  we consider the problem of  one-dimensional harmonic oscillator and the relativistic Hydrogen  and we solve
equations (\ref{a32}) or (\ref{a33}) to find  their energy spectrum.
\section{Relativistic Micro-Harmonic Oscillator}
The relativistic Generalization of the classical or quantum harmonic oscillator has been
debated  in literature (see, e. g., \cite{Guerrero, Harvey}). Yet, there has been provided no unique definition of the
relativistic harmonic oscillator, at least in micro-domain \cite{Cotaescu}. The main difficulty
is that using a Lorentz-invariant variational principle in classical domain, one can show that
the mass is potential-dependent \cite{Harvey}. So, the total energy of the system should
be expressed as (\ref{a32}). Strictly speaking, the relativistic energy of the particle can be
defined as:
\begin{equation}\label{b1}
E_{rP}=\gamma_{P}\left(\frac{1}{2}m_{0}w_{0}^{2}x^{2}+m_{0}c^{2}\right)
\end{equation}
where $\gamma_{P}$ is defined in (\ref{a23}) and $w_{0}$ is the spring-frequency of the
oscillator. Since the potential energy of the particle includes the relativistic mass
(and so is a function of the velocity of particle), its definition as a Hermitian operator
faces a difficulty. Thus, in perturbation methods as well as the Klein-Gordon solutions
of the relativistic quantum harmonic oscillator, the potential-dependency of mass is
usually ignored and similar to the relation (\ref{a33}), the relativistic energy of the particle
is considered as
\begin{equation}\label{b2}
E_{rP}=\frac{1}{2}m_{0}w_{0}^{2}x^{2}+\gamma_{P}m_{0}c^{2}.
\end{equation}
Using Lagrangian approach leading to the Klein-Gordon equation of a harmonic oscillator
in which the potential of the oscillator is assumed to be nonrelativistic, the
relativistic energy of the system can be obtained as \cite{Harvey}:
\begin{eqnarray}\label{n3}
E_{n}&=&\pm m_{0}c^{2}(1+\frac{2E^{0}_{n}}{m_{0}c^{2}})^{\frac{1}{2}}\\
&=&\pm (m_{0}c^{2}+E^{0}_{n}-\frac{{E^{0}_{n}}^{2}}{2m_{0}c^{2}}+...)
\end{eqnarray}
where $E^{0}_{n}=\hbar w_{0}(n+\frac{1}{2})$. Here, we see that $E^{0}$ appears as
the first order correction of $E_{n}$, not as a correction-free term.\\
\indent
Now, we examine the equation (\ref{a32}) and (\ref{a33}) for solving the same problem, using the relations
(\ref{b1}) and (\ref{b2}), respectively. To solve the equation (\ref{a32}), we first use the following
approximation, having attention that for a simple harmonic oscillator we can ignore
higher powers of $x$ (power greater than 2) in the potential expression. Hence,
\begin{equation}
(1+\frac{V_{nrP}}{m_{0}c^{2}})^{-2}\simeq (1-\frac{2V_{nrP}}{m_{0}c^{2}})
\end{equation}
where $V_{nrP}=\frac{1}{2}m_{0}w^{2}_{0}x^{2}$. Then, the equation (\ref{a32}) can be rearranged  as:
\begin{equation}\label{b3}
\chi^{\prime \prime}(x)+(\beta_{1}-\alpha_{1}^{2}x^{2})\chi(x)=0
\end{equation}
where
\begin{equation}
\beta_{1}= \frac{E^{2}-m_{0}^{2}c^{4}}{\hbar^{2}c^{2}};\hspace{1cm}\alpha_{1}^{2}=\frac{E^{2}w_{0}^{2}}{\hbar^{2}c^{4}}
\end{equation}
The equation (\ref{b3}) has precisely the form of the time-independent Schr\"{o}dinger
equation for nonrelativistic harmonic oscillator. So, one at once can obtain the energy
of the relativistic harmonic PF system. Here, we have:
\begin{equation}\label{n2}
2\alpha_{1}(n+\frac{1}{2})=\beta_{1}
\end{equation}
leading to:
\begin{equation}
E^{2}-2EE_{n}^{0}-m_{0}^{2}c^{4}=0,
\end{equation}
thus, we obtain:
\begin{equation}
E_{n}=E_{n}^{0} \pm m_{0}c^{2}(1+\frac{{E^{0}_{n}}^{2}}{m_{0}^{2}c^{4}})^{\frac{1}{2}}.
\end{equation}
Choosing the plus sign, the energy $E_{n}$ can be expanded as
\begin{equation}
E_{n}=E_{n}^{0}+ m_{0}c^{2}+\frac{{E^{0}_{n}}^{2}}{2m_{0}c^{2}}+...
\end{equation}
This result has not been recorded in literature so far. The normalized
eigenfunctions (corresponding to the stationary fields  $\chi_{n}(x)$ in (80)) can be written as :
\begin{equation}
\psi_{n}(x)=(d^{n}n!)^{-\frac{1}{2}}(\frac{\alpha_{1n}}{\pi})^{\frac{1}{4}}
\exp(\frac{\alpha_{1n}x^{2}}{2})H_{n}(\sqrt{\alpha_{1n}x})
\end{equation}
where $H_{n}(\sqrt{\alpha_{1n}x})$ are the Hermite polynomials. Assuming that $E_{n}^{0}\ll m_{0}c^{2}$
and considering $E_{n}\approx m_{0}c^{2}$ in $\alpha_{1n}$, we get
 $\alpha_{1}\approx\frac{m_{0}w_{0}}{\hbar}$ like its nonrelativistic definition. \\
 \indent
 On the other hand, taking in to account the relation (\ref{b2}), we can write
 the equation (\ref{a33}) as:
 \begin{equation}\label{n1}
 \chi^{\prime \prime}(x)+(\beta_{2}-\alpha_{2}^{2}x^{2})\chi(x)=0
 \end{equation}
 where
 \begin{equation}
\beta_{2}= \beta_{1}=\frac{E^{2}-m_{0}^{2}c^{4}}{\hbar^{2}c^{2}};\hspace{1cm}\alpha_{2}^{2}=\frac{E m_{0}w_{0}^{2}}{\hbar^{2}c^{2}}
 \end{equation}
 \indent
 In reaching the equation (\ref{n1}), we have assumed that $(E-V_{nrp})^{2}\simeq(E^{2}-2EV_{nrp})$. The coefficients $\alpha_{2}$ and $\beta_{2}$ satisfy the similar
 relation as (\ref{n2}). Accordingly, we get:
\begin{equation}
E^{2}-2\frac{c}{w_{0}}(\sqrt{m_{0}w_{0}^{2}E})E_{n}^{0}-m_{0}^{2}c^{4}=0
\end{equation}
\indent
To solve the above equation, we assume that $E_{n}^{0}\ll m_{0}^{2}c^{2}$, so that one can put inside the radical $E\approx m_{0}^{2}c^{2}$. Then, we obtain:
\begin{equation}
E^{2}=m_{0}^{2}c^{4}(1+\frac{2E_{n}^{0}}{m_{0}c^{2}})
\end{equation}
from which the same solution as (\ref{n3}) is resulted. In effect, the solution of our relativistic Schrodinger equation (\ref{a33}) for the case of nonrelativistic
harmonic potential coincides with the answer obtained by Lagrangian approach leading to the Klein-Gordon equation.
\section{Relativistic Schr\"{o}dinger Equation For  Hydrogen Atom}
The Hydrogen atom consists of a heavy, essential motionless proton (we may as well put it at the origin), of charge e, together with a much lighter electron
(charge -e) that orbits around it, bound by the mutual attraction of opposite charges. From Coulomb's law, the potential energy (in SI units) is\\
\begin{equation}\label{r2}
V(r)=-\frac{1}{4\pi‎\varepsilon‎‎_{0}}\frac{e^{2}}{r}.
\end{equation}
%
%
%
Since the potential of Hydrogen atom is dependent of mass, the field, $\chi$, and the energy of the PF system, $E_{PF}$, satisfies relation (\ref{a33}).
Our problem is to solve this equation for $\chi$, and determine the allowed energies, $E$.

The generalization to three dimension of relation (\ref{a33}) is straightforward. Typically, the potential is a function only of the distance from the origin.
In that case it is natural to adopt spherical coordinates  $(r, \theta, \phi)$. In spherical coordinates
 the relation  (\ref{a33})  takes the form
\begin{eqnarray}\label{r3}
&-&\frac{\hbar^{2}}{2m_{0}}\frac{1}{r^{2}}\left[ \frac{\partial}{\partial r}(r^{2}\frac{\partial}{\partial r}) +\frac{1}{\sin\theta}\left(\frac{\partial}{\partial \theta}(\sin\theta\frac{\partial}{\partial \theta})
+\frac{1}{\sin\theta}\frac{\partial^{2}}{\partial \phi^{2}}\right)\right]\chi({\bf r}) \nonumber \\
&+&\frac{1}{2}m_{0}c^{2}\chi({\bf r})=\frac{1}{2m_{0}c^{2}}(E-V_{nrp}(r))^{2}\chi({\bf r}).
\end{eqnarray}
In relation (\ref{r3}),  the potential  $V_{nrp}(r)$  is the classical potential defined by  (\ref{r2}).
 We begin by looking for solutions that are factorable in $r$ and angular variables $\theta$ and $\phi$:
\begin{equation*}
\chi({\bf r})=R(r)Y(\theta,\phi).
\end{equation*}
Putting this into relation  (\ref{r3}), we have
\begin{eqnarray}
&-&\frac{\hbar^{2}}{2m_{0}}Y(\theta,\phi)\frac{1}{r^{2}}\frac{d}{dr}(r^{2}\frac{d}{d r})R(r)\nonumber \\
&-&\frac{\hbar^{2}}{2m_{0}}\frac{R(r)}{r^{2}\sin\theta}\left(\frac{\partial}{\partial \theta}(\sin\theta\frac{\partial}{\partial \theta})
+\frac{1}{\sin\theta}\frac{\partial^{2}}{\partial\phi^{2}}\right)Y(\theta,\phi) \nonumber \\
&+&\frac{1}{2}m_{0}c^{2}R(r)Y(\theta,\phi)
=\frac{1}{2m_{0}c^{2}}(E-V_{nrp})^{2}R(r)Y(\theta,\phi).
\end{eqnarray}
Dividing by
$R(r)Y(\theta,\phi)$
and multiplying by  $-\frac{2m_{0}r^{2}}{\hbar^{2}}$, one gets:
\begin{eqnarray}
[\frac{1}{R(r)} \frac{d}{d r}(r^{2}\frac{d‎R(r)}{d r})+\frac{r^{2}}{\hbar^{2}c^{2}}(E-V_{nrp}(r))^{2}-\frac{r^{2}}{\hbar^{2}}m_{0}^{2}c^{2}] \nonumber\\
+[\frac{1}{Y(\theta,\phi)}\frac{1}{\sin\theta}\left(\frac{\partial}{\partial \theta}\sin\theta\frac{\partial}{\partial\theta}
+\frac{1}{\sin\theta}\frac{\partial^{2}}{\partial\phi^{2}}\right)Y(\theta,\phi)]=0.
\end{eqnarray}
The term in the first curly bracket depends only on $r$, where as the reminder depends only on $\theta$ and $\phi$. Accordingly, each must be a constant.
 We will write separation constant in the form  $l(l+1)$ \cite{Griffits},
\begin{eqnarray}
\frac{1}{R(r)} \frac{d}{d r}(r^{2}\frac{d‎R(r)}{d r})+\frac{r^{2}}{\hbar^{2}c^{2}}(E-V_{nrp}(r))^{2}-\frac{r^{2}}{\hbar^{2}}m_{0}^{2}c^{2}=l(l+1) \label{r107}\\
\frac{1}{Y(\theta,\phi)}\frac{1}{\sin\theta}\left(\frac{\partial}{\partial \theta}\sin\theta\frac{\partial}{\partial\theta}
+\frac{1}{\sin\theta}\frac{\partial^{2}}{\partial\phi^{2}}\right)Y(\theta,\phi)=-l(l+1) \label{r207}
\end{eqnarray}
The angular part of the wave function, $Y_{l}^{m}(\theta,\phi)$, is the same for all spherically symmetric potentials and it has been solved in
 most of text books  of quantum mechanics. So, the solution of the angular part of Hydrogen atom given by (\ref{r207}) is
$Y_{l}^{m}(\theta,\phi)$, that are called spherical harmonic \cite{Griffits}.
 On the other hand the actual shape of the potential, $V(r)$ affects only the radial part of the wave function R(r)  which
 is determined by relation (\ref{r107}). We rewrite it as
\begin{equation}
 \frac{d}{d r}(r^{2}\frac{d R(r)}{d r})+\frac{r^{2}}{\hbar^{2}c^{2}}(E^{2}-2EV+V^{2})R(r)-\frac{r^{2}}{\hbar^{2}}m_{0}^{2}c^{2}R(r)=l(l+1)R(r),
 \end{equation}
 or
 \begin{equation}\label{r9}
 \frac{d}{d r}(r^{2}\frac{d R(r)}{d r})+\frac{r^{2}}{\hbar^{2}c^{2}}(E^{2}-m_{0}^{2}c^{4}-2EV+V^{2})R(r)=l(l+1)R(r).
\end{equation}
Let
\begin{equation}\label{r10}
u(r)=rR(r),
\end{equation}
 so one can rewrite the relation (\ref{r9})  as
\begin{equation}\label{r11}
\frac{d^{2}u(r)}{d r^{2}}+\frac{1}{\hbar^{2}c^{2}}(E^{2}-m_{0}^{2}c^{4}-2EV+V^{2})u(r)=\frac{l(l+1)}{r^{2}}u(r).
\end{equation}
Putting the relation (\ref{r2}) in  (\ref{r11}), we find
\begin{eqnarray}\label{r7}
\frac{d^{2}u(r)}{d r^{2}}+[\frac{1}{\hbar^{2}c^{2}}(E^{2}-m_{0}^{2}c^{4})+\frac{2E}{\hbar^{2}c^{2}}\frac{e^{2}}{4\pi\varepsilon_{0}}\frac{1}{r} \nonumber \\
+(\frac{1}{\hbar^{2}c^{2}}(\frac{e^{2}}{4\pi‎\varepsilon‎‎_{0}})^{2}-l(l+1))\frac{1}{r^{2}}]u(r)=0
\end{eqnarray}
For simplifying the form of the relations,  we write relation  (\ref{r7})  as
\begin{equation}\label{rr1}
\frac{d^{2}u}{dr^{2}}=\left[-\frac{A}{r}+\frac{B(B+1)}{r^{2}}+D^{2}\right]u,
\end{equation}
which $A$, $G$, $B$ and $D$ are defined as:
\begin{subequations}
\begin{eqnarray}
A&=&\frac{2E}{\hbar^{2}c^{2}}\frac{e^{2}}{4\pi‎\varepsilon_{0}}, \label{a200}\\
G^{2}&=&(\frac{e^{2}}{4\pi‎\varepsilon_{0}})^{2}\frac{1}{\hbar^{2}c^{2}}, \label{b200}\\
B(B+1)&=&l(l+1)-G^{2}, \label{c200}\\
D^{2}&=&\frac{1}{\hbar^{2}c^{2}}\left(m_{0}^{2}c^{4}-E^{2}\right). \label{d200}
\end{eqnarray}
\end{subequations}
Since $m_{0}^{2}c^{4}\gg E^{2}$,  we always have $D^{2}>0$. We introduce
$\rho=Dr$  and  $\frac{A}{D}=\rho_{0} $,  so the equation (\ref{rr1}) is derived as
\begin{equation}\label{11}
\frac{d^{2}u}{d\rho^{2}}=\left[1-\frac{\rho_{0}}{\rho}+\frac{B(B+1)}{\rho^{2}}\right]u.
\end{equation}
This equation can be solved like the way the  radial part of the Schr\"{o}dinger equation of Hydrogen atom is solved.  We introduce the new function
  $v(\rho) $ as:
\begin{equation}\label{r19}
u(\rho)=\rho^{B+1}e^{-\rho}v(\rho).
\end{equation}
In terms of  $v(\rho)$,  the relation  (\ref{11})  reads as
\begin{equation}\label{r13}
\rho\frac{d^{2}v}{d\rho^{2}}+2(B+1-\rho)\frac{dv}{d\rho}+[\rho_{0}-2(B+1)v]=0.
\end{equation}
Finally, we assume the solution  $v(\rho)$  can be expressed as a power series in $\rho$:
\begin{equation}
v(\rho)=\sum c_{j}\rho^{j}
\end{equation}
Now, our problem  is  to determine the coefficient   $c_{j}$.
Inserting this in to the equation  (\ref{r13}), we have
\begin{equation}\label{r22}
c_{j+1}=\left[\frac{2(j+B+1)-\rho_{0}}{(j+1)(j+2B+2)}\right]c_{j}.
\end{equation}
This recursion formula determines  the coefficients, and hence the function $v(\rho)$. Fore large values of  $j$,
 the solutions aren't normalizable. To get rid of this dilemma, the series must terminate. There must occurs some maximal integer,
 $j_{max}$, such that
\begin{equation*}
c_{j_{max}+1}=0.
\end{equation*}
Evidently
\begin{equation*}
2(j_{max}+B+1)-\rho_{0}=0
\end{equation*}
and
\begin{equation*}
\rho_{0}=2(j_{max}+B+1).
\end{equation*}
 Using the relation  (\ref{c200}) we find
\begin{equation*}
B^{2}+B-l(l+1)+G^{2}=0,
\end{equation*}
So,
\begin{eqnarray}
B&=&\frac{1}{2}\left(-1\pm\sqrt{1-4[-l(l+1)+G^{2}]}\right) \nonumber \\
&=&-\frac{1}{2}+\sqrt{(l+\frac{1}{2})^{2}-G^{2}},
\end{eqnarray}
and  $\rho_{0}$ is obtained as
\begin{equation}
\rho_{0}=2\left(j_{max}+\frac{1}{2}+\sqrt{(l+\frac{1}{2})^{2}-G^{2}}\right).
\end{equation}
Then from the relation (\ref{r22}) we have
\begin{equation}\label{rr25}
c_{j+1}=\left[\frac{2(j+\frac{1}{2}+\sqrt{(l+\frac{1}{2})^{2}-G^{2}})-\rho_{0}}{(j+1)(j+2\sqrt{(l+\frac{1}{2})^{2}-G^{2}})+1)}\right]c_{j}.
\end{equation}
Now, with substituting  $A$ and $D$, from the relations (\ref{a200}) and (\ref{b200}) and having  $\rho_{0}=\frac{A}{D}$,
one obtains:
\begin{eqnarray}
\rho_{0}=\frac{A}{D}&=&2\left(j_{max}+\frac{1}{2}+\sqrt{(l+\frac{1}{2})^{2}-G^{2}}\right)  \nonumber \\
&=&\frac{\frac{2E}{\hbar^{2}c^{2}}\frac{e^{2}}{4\pi‎\varepsilon_{0}}}{\sqrt{\frac{1}{\hbar^{2}c^{2}}\left(m_{0}^{2}c^{4}-E^{2}\right)}}. \nonumber
\end{eqnarray}
So, with a little calculation, we can obtain  the value of $E$ as,
\begin{equation}\label{4r28}
E=\pm\frac{m_{0}c^{2}\left(j_{max}+\frac{1}{2}+\sqrt{(l+\frac{1}{2})^{2}-G^{2}}\right)}{\sqrt{G^{2}+\left(j_{max}+\frac{1}{2}+\sqrt{(l+\frac{1}{2})^{2}
-G^{2}}\right)^{2}}}.
\end{equation}
The relation (\ref{4r28}) shows the energy  of the PF system related to Hydrogen atom.
Using the relation (\ref{b200}), the value of  $G$  is obtained from:
$$G^{2}=(\frac{e^{2}}{4\pi\varepsilon_{0}})^{2}\frac{1}{\hbar^{2}c^{2}}\simeq5.4\times10^{-5}.$$
Since $\frac{G^{2}}{(l+\frac{1}{2})^{2}}\ll 1$, we can simplify the relation (\ref{4r28}) as follows.
Considering
\begin{eqnarray}\label{3r28}
\left[(l+\frac{1}{2})^{2}-G^{2}\right]^{\frac{1}{2}}&=&\left[(l+\frac{1}{2})^{2}\left(1-\frac{G^{2}}{(l+\frac{1}{2})^{2}}\right)\right]^{\frac{1}{2}}\nonumber\\
&\simeq&(l+\frac{1}{2})-\frac{G^{2}}{2(l+\frac{1}{2})}-\frac{1}{8}\frac{G^{4}}{(l+\frac{1}{2})^{3}},
\end{eqnarray}
%
and
\begin{eqnarray}\label{2r28}
\left[G^{2}+\left(j^{\prime}+l^{\prime}\sqrt{(1-\frac{G^{2}}{l^{\prime 2}}}\right)^{2}\right]^{-\frac{1}{2}}
&=&\left[ j^{\prime 2} + l^{\prime 2}+ 2 j^{\prime}  l^{\prime } \sqrt{1-\frac{G^{2}}{l^{\prime 2}}} \right]^{-\frac{1}{2}} \nonumber \\
&\simeq&( j^{\prime }+ l^{\prime })^{-1}\left[1-\frac{ j^{\prime}  l^{\prime }}{( j^{\prime }+ l^{\prime })^{2}}[\frac{G^{2}}{l^{\prime 2}}+\frac{1}{4}(\frac{G^{2}}{l^{\prime 2}})^{2}+...]\right]^{-\frac{1}{2}}\nonumber \\
&\simeq&\frac{1}{n}\left[  1+\frac{ j^{\prime}  l^{\prime }}{n^{2}}\frac{G^{2}}{2l^{\prime 2}} + \frac{1}{8}\frac{ j^{\prime}  l^{\prime }}{n^{2}}(1+3\frac{ j^{\prime}  l^{\prime }}{n^{2}})(\frac{G^{2}}{l^{\prime 2}})^{2}+... \right]
\end{eqnarray}
where we have ignored the higher power of $G^{4}$.  $n, j^{\prime }$ and $ l^{\prime } $  are defined as:
\begin{eqnarray}\label{1r28}
n&=&j_{max}+l+1,\nonumber \\
 j^{\prime }&=&j_{max}+\frac{1}{2},\nonumber \\
 l^{\prime }&=&l+\frac{1}{2}
\end{eqnarray}
So, putting relations (\ref{1r28}), (\ref{2r28}) and (\ref{3r28}) in  (\ref{4r28}), we have:
\begin{eqnarray}\label{r28}
E&=&m_{0}c^{2}[1-\frac{(l+\frac{1}{2})^{2}}{2n^{2}}\frac{G^{2}}{(l+\frac{1}{2})^{2}}+\frac{48(l+\frac{1}{2})^{4}-64n(l+\frac{1}{2})^{3}}{128n^{4}}(\frac{G^{2}}{(l+\frac{1}{2})^{2}})^{2}]\nonumber \\
&=&m_{0}c^{2}[1-\frac{G^{2}}{2n^{2}}+\frac{48G^{4}}{128n^{4}}-\frac{64G^{4}}{128n^{3}(l+\frac{1}{2})}].\nonumber \\
\end{eqnarray}
%
 Eliminating  $G$ by   using  the relation  (\ref{b200}), and having into account that
$E_{n}=-\frac{m_{0}}{2\hbar^{2}}(\frac{e^{2}}{4\pi\varepsilon_{0}})^{2}\frac{1}{n^{2}}$, it follows:
%
%
\begin{equation}\label{5r28}
E=m_{0}c^{2}+E_{n}-\frac{E_{n}^{2}}{2 m_{0}c^{2}}[\frac{4n}{(l+\frac{1}{2})}-3]
\end{equation}
As we see, the relation (\ref{5r28})  is exactly the same as relativistic correction to the energy levels of Hydrogen atom
obtained by the first-order time independent perturbation theory. \\
%
\indent 
Finally, the spatial wave function labeled by three quantum numbers, $n$, $l$ and $m$
is
\begin{equation}
\psi_{nlm}(r, \theta, \phi)=R_{nl}(r)Y^{m}_{l}(\theta, \phi),
\end{equation}
where, referring back to  (\ref{r10}) and (\ref{r19}) for $R(r)$,
\begin{equation}
R_{nl}(r)=\frac{1}{r}\rho^{B+1}e^{-\rho}v(\rho)
\end{equation}
in which $v(\rho)$ is a polynomial of degree $j_{max}$  whose coefficients are determined by the recursion
formula, (\ref{rr25}).
\section{Conclusion}

  It is indeed one of the main successes of our theory to provide with us a coherent way for describing conservative systems. In \cite{shafiee4} we derived the relativistic Schr\"{o}dinger equation for the case that the potential energy of the particle includes the relativistic mass and the case that it is independent of it.  In this essay, we solved the relativistic Schr\"{o}dinger equation to find the energy spectrum of a relativistic micro-harmonic oscillator and relativistic Hydrogen atom.\\
\indent
   Considering the relativistic micro-harmonic oscillator, we examined relations (\ref{a32}) and (\ref{a33}) for solving the same problem, using the relations (\ref{b1}) and (\ref{b2}), respectively. The solution of the relativistic Schr\"{o}dinger equation (\ref{a33}) for the case of nonrelativistic harmonic potential coincides with the answer obtained by Lagrangian approach leading to the Klein-Gordon equation.\\
\indent
Moreover,  we  found the energy spectrum of relativistic Hydrogen atom using the relativistic Schr\"{o}dinger equation for potential energy that is independent of mass.
It is one of our achievements that the result of energy spectrum in (\ref{5r28}) for relativistic Hydrogen atom is completely consistent with the  relativistic correction to the energy levels of Hydrogen atom obtained by the first-order time independent perturbation theory.\\

\end{document}